\documentclass[prl,twocolumn,showpacs,lengthcheck,superscriptaddress]{revtex4}

\usepackage{amsmath}
\usepackage{amssymb}
\usepackage{amsthm}
\usepackage{graphicx}

\newcommand{\tr}{\operatorname{tr}}
\newcommand{\uinvnorm}{|\kern-2pt|\kern-2pt|}

\begin{document}
\bibliographystyle{apsrev}

\title{General Monogamy Inequality for Bipartite Qubit Entanglement}

\author{Tobias J.\ Osborne}
\email[]{T.J.Osborne@bristol.ac.uk}
\affiliation{Department of
Mathematics, University of Bristol, University Walk, Bristol BS8
1TW, United Kingdom}
\author{Frank Verstraete}
\email[]{fverstraete@ist.caltech.edu} \affiliation{Institute for
Quantum Information, California Institute of Technology, Pasadena,
CA 91125, USA}

\date{\today}

\begin{abstract}
We consider multipartite states of qubits and prove that their
bipartite quantum entanglement, as quantified by the concurrence,
satisfies a monogamy inequality conjectured by Coffman, Kundu, and
Wootters. We relate this monogamy inequality to the concept of
frustration of correlations in quantum spin systems.

\end{abstract}


\maketitle

Quantum mechanics, unlike classical mechanics, allows the
existence of \emph{pure states} of composite systems for which it
is not possible to assign a definite state to two or more
subsystems. States with this property are known as \emph{entangled
states}.  Entangled states have a number of remarkable features, a
fact which has inspired an enormous literature in the years since
their discovery. These properties have led to suggestions that the
propensity of multipartite quantum systems to enter nonlocal
superposition-states might be the \emph{defining} characteristic
of quantum mechanics \cite{schrodinger:1935a, bell:1964a}.

It is becoming clear that entanglement is a physical resource. The
exploration of this idea is a central goal in the burgeoning field
of quantum information theory. As a consequence, the study of the
mathematics underlying entanglement has been a very active area and
has led to many operational and information-theoretic insights. As
for now, only the pure-state case of entanglement shared between two
parties is thoroughly understood and quantified; progress on the
multipartite setting has been much slower.

A key property, which maybe as fundamental as the no-cloning
theorem, has been discovered recently in the context of multipartite
entanglement: \emph{entanglement is monogamous} \cite{coffman:2000a,
terhal:2003a}. More precisely, there is an inevitable tradeoff
between the amount of quantum entanglement that two qubits $A$ and
$B$, in Alice's and Bob's possession, respectively, can share and
the quantum correlation that Alice's same qubit $A$ can share with
Charlie, a third party, $C$ \cite{coffman:2000a}. In the context of
quantum cryptography, such a monogamy property is of fundamental
importance because it quantifies how much information an
eavesdropper could potentially obtain about the secret key to be
extracted. The constraints on shareability of entanglement lie also
at the heart of the success of many information-theoretic protocols,
such as entanglement distillation.

In the context of condensed matter physics, the monogamy property
gives rise to the frustration effects observed in, e.g., Heisenberg
antiferromagnets. Indeed, the perfect ground state for an
antiferromagnet would consist of singlets between all interacting
spins. But, as a particle can only share one unit of entanglement
with all its neighbours (this immediately follows from the dimension
of its local Hilbert space), it will try to spread its entanglement
in an optimal way with all its neighbours leading to a strongly
correlated ground state. The tools developed in this Letter will
allow us to turn such qualitative statements into quantitative ones.

The problem of fully quantifying the constraints on distributed
entanglement should be seen as analogous to the
\emph{N-representability problem} for fermions \cite{coleman:2000a}.
This is because, just as is the case for fermions, if the
constraints on distributed entanglement were known explicitly then
this would render trivial \cite{verstraete:2005a} the task of
computing the ground-state energy of condensed-matter systems. The
results of this Letter represent the first step towards the full
quantification of the constraints on distributed entanglement.


The main result of this Letter is a proof of the longstanding
conjecture of Coffman, Kundu, and Wootters \cite{coffman:2000a} that
the distribution of bipartite quantum entanglement, as measured by
the tangle $\tau$, amongst $n$ qubits satisfies a tight inequality:
\begin{equation}\label{eq:ckw}
\tau(\rho_{A_1A_2})+\tau(\rho_{A_1A_3})+ \cdots +
\tau(\rho_{A_1A_{n}}) \le \tau(\rho_{A_1(A_2A_3\cdots A_{n})}),
\end{equation}
where $\tau(\rho_{A_1(A_2A_3\cdots A_{n})})$ denotes the bipartite
quantum entanglement measured by the tangle across the bipartition
$A_1:A_2A_3\cdots A_n$. This inequality (which we shall henceforth
refer to as the \emph{CKW inequality}) has been established in the
case of three qubits. However, the case of $n$ qubits was still open
\cite{endnote24}. In this Letter we establish Eq.~(\ref{eq:ckw}) for
arbitrary numbers $n$ of qubits.

The outline of this Letter is as follows. We begin by introducing
and defining the quantum correlation measures we study throughout
this Letter. Following this we reduce the CKW inequality to a
statement pertaining to quantum correlation measures for a pure
tripartite system consisting of two qubits and a four-level quantum
system. Such a system is, up to local unitaries, completely
determined by its two-qubit reduced density operator. The proof will
then be completed by showing that the one-way correlation measure
\cite{henderson:2001a,koashi:2004a} of a mixed state of two qubits
is always larger or equal than its tangle.

In our proof we have utilised a number of techniques. We derive a computable formula for the linear Holevo $\chi$
quantity for all qubit maps and also for the one-way correlation measure \cite{henderson:2001a} for all two-qubit
states.

To quantify mixed-state bipartite quantum correlations we study two measures. We also study one channel capacity
measure. The first measure we consider is the \emph{tangle} $\tau(\rho_{AB})$ which is the square of the
\emph{concurrence} \cite{bennett:1996a, hill:1997a, wootters:1998a,coffman:2000a,osborne:2002b},
$\tau(\rho_{AB})= C^2(\rho_{AB})$. The tangle measure pertains to bipartite quantum states $\rho_{AB}$ of a qubit
$A$ and a $D$-level quantum sytem $B$. To define the tangle we introduce the following entropic measure, the
\emph{linear entropy} $S_2$, for single-qubit states $\rho$ \cite{endnote19}:
\begin{align*}
S_2(\rho) &\triangleq 2(1-\tr(\rho^2)) \\
&= 4\det(\rho).
\end{align*}
The linear entropy $S_2$ is concave and unitarily invariant.

The tangle $\tau$ is now defined for any state $\rho_{AB}$ of the
$2\times D$ system via the roof construction (for operational
motivations and further discussion of this construction see
\cite{bennett:1996a} and \cite{uhlmann:1998a})
\begin{equation}
\tau(\rho_{AB}) \triangleq \inf_{\{p_x, \psi_x\}} \sum_{x} p_x
S_2(\tr_{B}(\psi_x)),
\end{equation}
where the infimum runs over all pure-state decompositions $\{p_x,
\psi_x\}$ of $\rho_{AB}$, $\rho_{AB}= \sum_{x} p_x \psi_x$.

The second correlation measure we need is closely related to a one-way correlation measure
\cite{henderson:2001a,koashi:2004a}. For any mixed state $\rho_{AB}$ of a $2\times D$ bipartite quantum system we
define
\begin{equation}\label{eq:hvtau}
I^{\leftarrow}_2(\rho_{AB}) \triangleq
\max_{\{M_x\}}\left(S_2(\rho_A) - \sum_{x} p_x S_2(\rho_x)\right),
\end{equation}
where the maximum runs over all POVMs $\{M_x\}$ on Bob's system,
$p_x = \tr(I_B\otimes M_x \rho_{AB})$ is the probability of
outcome $x$, and $\rho_x = \tr_{B}(I_B\otimes M_x \rho_{AB})/p_x$
is the posterior state in Alice's subsystem.

The third measure we will need, the \emph{linear Holevo $\chi$}
capacity, is a capacity measure for qubit channels $\Lambda$. This
measure is related to the one-shot Holevo $\chi$ quantity and is
defined by
\begin{equation}\label{eq:holevochi}
\chi_2(\rho; \Lambda) = \max_{\{p_x, \psi_x\}} \left(
S_2(\Lambda(\rho)) - \sum_{x} p_x S_2(\Lambda(\psi_x)) \right),
\end{equation}
where $\rho$ is a qubit ensemble, $\Lambda$ is an arbitrary qubit
channel (a trace-preserving completely-positive map), and the
maximum runs over all pure state decompositions $\{p_x, \psi_x\}$
of $\rho$, $\rho = \sum_{x} p_x \psi_x$.

We now turn to the CKW inequality. Our strategy for proving
Eq.~(\ref{eq:ckw}) will be to prove it for states $\rho_{ABC}$ of
two qubits $AB$, and a $2^{n-2}$-dimensional qudit $C$. The next
step we use is to proceed via induction by successively
partitioning the last qudit $C$ into two subsystems, a qubit
$C_1$, and a $2^{n-3}$-dimensional qudit $C_2$, and establishing
Eq.~(\ref{eq:ckw}) for the (typically mixed) state
$\rho_{AC_1C_2}$. Thus, the formula we will try to prove is the
following
\begin{equation}\label{eq:ertangle1}
\tau(\rho_{A(BC)})\ge\tau(\rho_{AB}) + \tau(\rho_{AC}),
\end{equation}
for arbitrary states $\rho$ of a $2\times2\times2^{n-2}$ system
$ABC$.

We begin by trying to prove Eq.~(\ref{eq:ertangle1}) for pure states. In this case we can use the local-unitary
invariance of $\tau(\rho_{AC})$ to rotate the basis of subsystem $C$ into the local Schmidt basis $|u_j\rangle$,
$j=1,\ldots, 4$, given by the eigenvectors of $\rho_C$. In this way we can regard the $2^{n-2}$-dimensional qudit
$C$ as an \emph{effective} $4$-dimensional qudit. Therefore, it is sufficient to establish
Eq.~(\ref{eq:ertangle1}) for a $2\times2\times4$ system $ABC$.

Supposing we have proved the inequality Eq.~(\ref{eq:ertangle1})
for pure states we can extend Eq.~(\ref{eq:ertangle1}) to mixed
states $\rho$. Consider the minimising decomposition
$\{p_x,|\psi_x\rangle\}$ for $\tau(\rho_{A(BC)})$, and apply the
inequality Eq.~(\ref{eq:ertangle1}) to each term,
\begin{equation}\label{eq:ineqmixed}
\begin{split}
\tau(\rho_{A(BC)}) &= \sum_{x} p_x \tau(\rho_{A(BC)}^x),\\
    &\ge \sum_{x} p_x (\tau(\rho_{AB}^x) +
    \tau(\rho_{AC}^x)), \\
    &\ge \tau(\rho_{AB}) + \tau(\rho_{AC}),
\end{split}
\end{equation}
where $\rho_{A(BC)}^x = |\psi_x\rangle\langle\psi_x|$, and we have
used the convexity of $\tau$ to arrive at the third line.

Now all that is required to establish the inequality
Eq.~(\ref{eq:ertangle1}) for an arbitrary system of $n$ qubits is
to successively apply Eq.~(\ref{eq:ertangle1}) to partitions of
$C$ according to the inductive recipe outlined above.  We
illustrate this procedure for pure states $\rho$ of four qubits
$ABC_1C_2$. Let $C = C_1 C_2$ be a combined pair of qubits and
apply Eq.~(\ref{eq:ertangle1}),
\begin{equation}
\begin{split}
\tau(\rho_{A(BC)}) &\ge \tau(\rho_{AB}) + \tau(\rho_{AC}), \\
    &\ge \tau(\rho_{AB}) + \tau(\rho_{AC_1}) +
    \tau(\rho_{AC_2}),
\end{split}
\end{equation}
where we have applied the mixed-state version of the inequality
Eq.~(\ref{eq:ertangle1}) in the second line. It is straightforward
to generalise this procedure to $n$ qubits.

We have now reduced the CKW inequality to an inequality for the tangle for pure states of a tripartite system
$ABC$ of two qubits $A$ and $B$ and a four-level system $C$. In the case of pure states, $\rho_{AB}$ and
$\rho_{AC}$ contain the same information (up to local unitaries); all possible POVM measurents at Bob's side
induce all possible pure state decompositions of $\rho_{AC}$,  and therefore the following monogamy relation
holds (see also Koashi and Winter \cite{koashi:2004a}):
\begin{equation}\label{eq:kw}
S_2(\rho_A) = \tau(\rho_{A(BC)}) = I_2^{\leftarrow} (\rho_{AB}) + \tau(\rho_{AC}).
\end{equation}

By comparing Eq.~(\ref{eq:ertangle1}) and Eq.~(\ref{eq:kw}) we see
that in order to establish Eq.~(\ref{eq:ertangle1}) it is
sufficient to establish the inequality
\begin{equation}\label{eq:tauinf}
\tau(\rho_{AB}) \le I_2^{\leftarrow} (\rho_{AB}),
\end{equation}
for all two-qubit states $\rho_{AB}$. As a first step toward
proving this inequality, we will now derive a computable formula
for $I_2^{\leftarrow} (\rho_{AB})$.

Any bipartite quantum state $\rho_{AB}$ may be written as
\begin{equation}\label{eq:statechannel}
\rho_{AB} = \Lambda_{\rho}\otimes I_B(|r_{B'B}\rangle\langle r_{B'B}|),
\end{equation}
where $|r_{B'B}\rangle$ is the symmetric two-qubit purification of the reduced density operator $\rho_{B}$ on an
auxiliary qubit system $B'$ and $\Lambda_\rho$ is a qubit channel from $B'$ to $A$. It can now readily be seen
that the one-way correlation measure $I_2^{\leftarrow} (\rho_{AB})$ is equal to the one-shot channel capacity
measure $\chi_2(\rho_{B}; \Lambda_\rho)$: all possible POVM measurements induce all convex decompositions of
$\rho_{B'}$.

The action of a qubit channel $\Lambda$ on a single qubit state
$\rho = \frac{I + \mathbf{r}\cdot\boldsymbol{\sigma}}{2}$, where
$\boldsymbol{\sigma}$ is the vector of Pauli operators, may be
written as
\begin{equation}
\Lambda(\rho) = \frac{I + (\mathbf{L}\mathbf{r} +
\mathbf{l})\cdot\boldsymbol{\sigma}}{2},
\end{equation}
where $\mathbf{L}$ is a $3\times 3$ real matrix and $\mathbf{l}$ is
a three dimensional vector. In this Pauli basis, the possible
decompositions of $\rho_B$ into pure states are represented by all
possible sets of probabilities $\{p_j\}$ and unit vectors
$\{\mathbf{r}_j\}$ for which $\sum_jp_j \mathbf{r}_j=\mathbf{r}_B$
where $\frac{I+\mathbf{r}_B\cdot\boldsymbol{\sigma}}{2}=\rho_B$. The
linear entropy $S_2$, written in terms of the Bloch vector
$\mathbf{r}$ of a two-qubit state, is given by $S_2(\frac{I +
\mathbf{r}\cdot\boldsymbol{\sigma}}{2}) = 1- |\mathbf{r}|^2$. In
this way we see that
\begin{equation}
Q(\mathbf{r}) = S_2\left(\Lambda\left(\frac{I + \mathbf{r}\cdot\boldsymbol{\sigma}}{2}\right)\right) = 1-
(\mathbf{L}\mathbf{r} + \mathbf{l})^T(\mathbf{L}\mathbf{r} + \mathbf{l}),
\end{equation}
which is a quadratic form in the Bloch vector $\mathbf{r}$.

Substituting $\mathbf{r}_j=\mathbf{r}_B+\mathbf{x}_j$, one can
easily check that the calculation of $\chi_2(\rho_{B};
\Lambda_\rho)$ reduces to determining $\{p_j,\mathbf{x}_j\}$
subject to the conditions $\sum_jp_j\mathbf{x}_j=\mathbf{0}$ and
$\|\mathbf{r}_B+\mathbf{x}_j\|=1$ maximizing
\begin{equation}\label{max1}
\max_{\{p_j,\mathbf{x}_j\}}\sum_jp_j
\mathbf{x}_j^T\mathbf{L}^T\mathbf{L}\mathbf{x}_j.
\end{equation}
Let us, without loss of generality, assume that
$\mathbf{L}^T\mathbf{L}$ is diagonal with diagonal elements
$\lambda^x\geq\lambda^y\geq\lambda^z$. The constraints
$\|\mathbf{r}_B+\mathbf{x}_j\|=1$ lead to the identities
\[\left(\mathbf{x}_j^x\right)^2=1-\|\mathbf{r}_B\|^2-2\mathbf{r}_B^T\mathbf{x}_j-\left(\mathbf{x}_j^y\right)^2-\left(\mathbf{x}_j^z\right)^2.\]
Substituting this into (\ref{max1}), we get
\begin{eqnarray*}\label{max2}
\chi_2(\rho_{B};
\Lambda_\rho)&=&\lambda_x(1-\|\mathbf{r}_B\|^2)+\\&&\hspace{-2cm}\max_{\{p_j,\mathbf{x}_j\}}\sum_jp_j\left((\lambda_y-\lambda_x)\left(\mathbf{x}_j^y\right)^2
+(\lambda_z-\lambda_x)\left(\mathbf{x}_j^z\right)^2\right).
\end{eqnarray*}
This expression is obviously maximised by choosing
$\mathbf{x}_j^z=\mathbf{x}_j^y=0$ for all $j$; the
$\mathbf{x}_j^x$ then have to correspond to the roots of the
equation $\|\mathbf{r}_B+\mathbf{x}_j\|=1$. There are exactly two
such roots, showing that the the maximum
$\lambda_x(1-\|\mathbf{r}_B\|^2)$ can be reached by an ensemble of
two states.

As $S_2(\rho_B)=1-\|\mathbf{r}_B\|^2$, we therefore obtain the
following computable expression for the linear Holevo $\chi$
capacity for qubit channels:
\begin{equation}\label{eq:chi2expr}
\chi_2(\rho_B; \Lambda) = \lambda_{\text{max}}(\mathbf{L}^T\mathbf{L})S_2(\rho_B).
\end{equation}
From this expression we also obtain an expression for $I^{\leftarrow}_2(\rho_{AB})$ via the correspondence
Eq.~(\ref{eq:statechannel}).

Now that we have a formula for $I^{\leftarrow}_2(\rho_{AB})$, we
want to prove that it is always larger than or equal to
$\tau(\rho_{AB})$. First of all, we note that a local filtering
operation of the form $\rho_{AB}'= \frac{(\mathbb{I}\otimes
B)\rho_{AB}(\mathbb{I}\otimes B)^\dag}{\tr((\mathbb{I}\otimes
B^\dag B)\rho_{AB})}$ leaves $\mathbf{L}$ invariant and transforms
\[S_2(\rho_B')=\frac{\det(B)^2}{\tr((\mathbb{I}\otimes B^\dag
B)\rho_{AB})^2}S_2(\rho_B).\] It happens that $I_2^{\leftarrow}$
transforms in exactly the same way as the tangle does
\cite{verstraete:2001b} (recalling that the  tangle is the square
of the concurrence). As there always exists a filtering operation
for which $\rho_B'\propto I_2$, we can assume, without loss of
generality, that $S_2(\rho_B)=1$.

So let's consider $\rho_{AB}$ with ${\rm Tr}_A(\rho_{AB})=
\frac{1}{2}I$. As
$\lambda_{\text{max}}(\mathbf{L}^T\mathbf{L})=\sigma^2_{\max}(\mathbf{L})$
where $\sigma_{\max}(\mathbf{L})$ is the largest singular value of
$\mathbf{L}$, we want to prove that $\sigma_{\max}(\mathbf{L})\geq
C(\rho_{AB})$ where $C(\rho_{AB})$ denotes the concurrence of
$\rho_{AB}$. It has been proven in \cite{verstraete:2002a} that any
mixed state of two qubits with associated $3\times 3$ matrix
$\mathbf{L}_{jk}={\rm Tr}(\rho\sigma_j\otimes\sigma_k)$ can be
written as a convex decomposition of rank-$2$ density operators all
having the same $\mathbf{L}_{jk}$. As the concurrence is convex, the
maximum concurrence for a given $\mathbf{L}$ will certainly be
achieved for a rank-$2$ density operator $\rho_2$. Next notice that
any rank-2 matrix $\rho_2$ can, up to local unitaries, be written as
\[\rho_2=p|00\rangle\langle 00|+(1-p)|\psi\rangle\langle\psi|.\]
Given the concurrence of $C(|\psi\rangle\langle\psi|)=C$, then
obviously $C(\rho_2)\leq (1-p)C$. Let us now consider
$\sigma_{\max}(\mathbf{L})$; this is the largest singular value of
the sum of two matrices, one having singular values $[p,0,0]$ and
the other one having $(1-p)[C,C,1]$ (corresponding to $|00\rangle$
and $|\psi\rangle$). Up to left and right multiplication by
unitaries, $\mathbf{L}$ is then given by
\[\mathbf{L}=(1-p)\left(\begin{array}{ccc}C&0&0\\0&C&0\\0&0&1\end{array}\right)+p\left(\begin{array}{c}\cos(\phi)\\0\\\sin(\phi)\end{array}\right)\mathbf{u}^T\]
where $\mathbf{u}$ is a unit vector. Obviously, the $(2,2)$ element of this matrix is $(1-p)C$, which is
certainly a lower bound for $\sigma_{\max}(\mathbf{L})$. This therefore implies that $I_2^{\leftarrow}(\rho_{AB})
\ge \tau(\rho_{AB})$ for all two-qubit states $\rho_{AB}$, hence proving the CKW inequality Eq.~(\ref{eq:ckw}).

The CKW inequality is likely to be useful in a number of contexts,
allowing simplified proofs of no-broadcasting bounds and
constraints for qubit multitap channel capacities. Perhaps the
most interesting open problem at this stage is to generalise
Eq.~(\ref{eq:ckw}) to systems other than qubits and to the case
where $A_1$ consists of more than one qubit. In both these cases
the available generalisations of the tangle measure for quantum
entanglement provably cannot yield entanglement sharing
inequalities. It is an interesting open problem to work out an
easily computable measure of quantum entanglement which will yield
concrete useful bounds on the distribution of private
correlations.

The CKW inequality may be immediately applied to study the
entanglement for a wide class of complex quantum systems. Let us,
for example, consider a translation-invariant state of a quantum
spin $1/2$ system on a lattice with coordination number $d$. The CKW
inequality implies that the concurrence $C(\rho)$ of the reduced
density operator $\rho$ of two nearest neighbours satisfies
$C(\rho)\leq (1-\langle S_\mathbf{n}\rangle^2)/\sqrt{d}$, where
$\langle S_\mathbf{n}\rangle$ is the magnetisation in the direction
$\mathbf{n}$. Hence the CKW inequality provides a quantitative tool
of assessing how far the mean-field energy will be from the exact
one. Let's e.g. consider the Heisenberg Hamiltonian. As the overlap
of a state $\rho$ with a singlet is bounded above by $(1+C(\rho))/2$
\cite{verstraete:2002a} and as the mean field energy per bond is
given by $1/2$, the gap between mean field theory and and the exact
ground state density \cite{dowling:2004a} is bounded above by
$(1-\langle S_\mathbf{n}\rangle^2)/(2\sqrt{d})$. The classical
result \cite{werner:1989a} that mean field theory becomes exact,
i.e.\ $\rho$ is separable, when $d\rightarrow\infty$ is a limiting
case of this inequality.

In a similar context, several investigations of the constraints on distributed entanglement have been carried out
recently. We mention, for example, \cite{roscilde:2004a, roscilde:2005a}. The validity of some results of these
papers were conditioned on the truth of the CKW inequality. As a consequence of this Letter it is now possible to
regard these results as true.

In this Letter we have proved that the distribution of bipartite
quantum entanglement is subject to certain shareability laws. It is
tempting to think that such shareability constraints might hold for
other quantum correlation quantities, such as the Bell violation of
a bipartite Bell inequality. This is in fact the case; it has
recently been discovered \cite{toner:2006a} that bipartite Bell
violations cannot be distributed arbitrarily.

Finally, it is worth highlighting some classes of quantum states
which saturate the CKW inequality. The classic example of a quantum
state saturating the CKW inequality is the $W$-state
\begin{equation*}
|W\rangle=\frac{1}{\sqrt{n}}\left(|0\cdots 01\rangle+|0\cdots
10\rangle+\cdots +|1\cdots 00\rangle\right).
\end{equation*}
The $W$-state has the property that the entanglement of any two
spins is equal, but the entanglement of the spin $A_1$ is not
maximal. One might ask if there are any states which saturate the
CKW inequality which have the property that the spin $A_1$ at the
focus is maximally entangled with the rest. In this way we could
regard such a state as sharing out a full unit of entanglement with
its neighbours. Such a state does indeed exist and is given by
\begin{equation*}
|\psi\rangle=\frac{1}{\sqrt{2}}|0\rangle|0\cdots
0\rangle+\frac{1}{\sqrt{2}}|1\rangle|W\rangle
\end{equation*}

In conclusion, we proved the Coffman-Kundu-Wootters monogamy
inequality which quantifies the frustration of entanglement between
different parties. The unique feature of this inequality is that it
is valid for any multipartite state of qubits, irrespective of the
underlying symmetries, which makes it much more general than de
Finetti type bounds \cite{koenig:2005a}. We also discussed the
relevance of the monogamous nature of entanglement in quantum
cryptography and in frustrated quantum spin systems.

\begin{acknowledgements}
We would particularly like to thank Michael Nielsen for
introducing TJO to this problem and for providing a tremendous
amount of encouragement and suggestions. Also, we are deeply
indebted to Andreas Winter for extensive helpful suggestions and
comments. We would also like to thank Bill Wootters for many
encouraging discussions. Finally, TJO is grateful to the EU for
support for this research under the IST project RESQ and also to
the UK EPSRC through the grant QIPIRC. FV is grateful to the
Gordon and Betty Moore Foundation.
\end{acknowledgements}


\begin{thebibliography}{22}
\bibitem{endnote19}{The Renyi $2$-entropy $R$ is often denoted as $S_2$
in the literature. While the linear entropy is equivalent to the
Renyi $2$-entropy, $R = -\protect \qopname  \relax o{log}(\protect
\operatorname  {tr}(\rho ^2))$, the reader should note that they are
not the same quantity.}
\end{thebibliography}
\end{document}